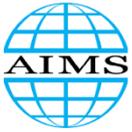

*Research article*

# A shiny app for modelling the lifetime in primary breast cancers patients through phase-type distributions

**Christian Acal[1], Elena Contreras[2], Ismael Montero[3] and Juan Eloy Ruiz-Castro[1,*]**

[1] Department of Statistics and O.R. and IMAG, University of Granada, 18071 Granada, Spain
[2] University of Granada, 18071 Granada, Spain
[3] Department of Statistics and O.R., University of Cádiz, 11510 Cádiz, Spain

**\* Correspondence:** jeloy@ugr.es; Tel: +34-958-246-306

**Abstract:** Phase-type distributions (PHD), which are defined as the distribution of the lifetime up to the absorption in an absorbent Markov chain, are an appropriate candidate to model the lifetime of any system, since any non-negative probability distribution can be approximated by a PHD with sufficient precision. Despite PHD potential, friendly statistical programs do not have implemented a module in their interfaces to handle PHD. Then, researchers must consider others statistical software such as R, Matlab or Python that work with the compilation of code chunks and functions. This fact might be an important handicap for those researchers who do not have sufficient knowledge in programming environments. In this paper, a new interactive web application developed with shiny is introduced in order to adjust PHD to an experimental dataset. This open access app does not require any kind of knowledge about programming or major mathematical concepts. Users can easily compare the graphic fit of several PHD while also estimating their parameters and assess the goodness of fit with just several clicks. All these functionalities are exhibited by means of a numerical simulation and modelling the time to live since the diagnostic in primary breast cancers patients.

**Keywords:** phase-type distributions; modelling; lifetime; shiny; interactive app; breast cancer

## 1. Introduction

In a wide sense, reliability analysis is a branch of statistics in charge of measuring the probability that a system performs correctly during a specific time duration, usually called 'lifetime' (or failure time). These systems are undergone a continuous wear by uncontrollable variables, so that each of them is going to fail (randomly) at different instants of time. In this context, the Probability Theory plays a fundamental role in the field of reliability analysis, since identifying the probability distribution of the lifetime might shed light on the existing variability behind the systems operation. Many of the traditional probability distributions such as Exponential, Weibull or Gamma have been employed for these purposes [1,2], even for modelling count data by constructing their discrete analogues [3]. However, due to the apparition of increasingly complex systems, these probability distributions display a pretty poor fit in many occasions. To solve this problem, a solution might be considering some kind of transformation on the data (see, for example, [4]) or applying a probability distributions mixture (see, for instance, [5,6]). Another suitable option is to consider an approach based on Phase-type distributions (PHD) [7,8].

PHD are defined as the distribution of the lifetime up to the absorption in an absorbent Markov chain with finite state space (for more information about Markov processes, see, for example, [9,10]). Its matrix-algebraic form makes possible to model complex systems with well-structured results, which facilitates the posterior interpretation. The main characteristics of PHD are reviewed in depth in [11]. An interesting result is any non-negative probability distribution may be approximated arbitrarily closely by a PHD, since this class of distributions is dense in the set of probability distributions on the non-negative half-line [12,13]. Thanks to these characteristics, PHD are highly considered in many areas of knowledge. Among other many contributions, PHD have shown a better behavior than Weibull or Exponential distributions to model data associated with non-volatile memories [14,15]; in [16] PHD are used to predict the length of stay in hospital for elderly patients; in [17] PHD are applied in the sector of risk theory; and a new class of distributions called Linear PHD was introduced to model the functional principal component analysis in [18].

The inherent problem that many researchers face when they are using PHD is that most friendly statistical software such as SPSS or Statgraphics do not have a module available to handle PHD. Needless to say of others software outside the area of Statistics. Then, researchers must resort to programmes such as R, Matlab or Python, in which it is necessary to run code chunks and collections of functions to obtain the outcomes. This fact may cause an important rejection due to the lack of knowledge in programming concepts and researchers would look for other alternatives, although the results are not good enough. Hence, there is a great need to develop new tools in which no kind of knowledge about programming or mathematical concepts are required to use them. In this way researchers could apply complex statistical analysis, no matter their area of research and education.

In this regard, many friendly and interactive applications are being built in recent years on R package shiny [19], which significantly reduces the barriers to producing webpage-style representations of analysis results in R. The interfaces produced by the shiny framework are rendered locally by a web browser. These apps can be hosted publicly and might be incorporated as important parts of reports and scientific papers [20]. In the literature, there are already a wide variety of shiny apps for different purposes. In [21] an interactive graphics for functional data analyses is introduced, while the app called *shinyMethyl* is aimed at the visualization of high-dimensional genomic data [22]. [23] built a shiny application that produced daily updates about the evolution of the SARS-CoV-2 epidemic worldwide. Within the field of reliability analysis, *shinystan* is developed for exploring Bayesian models fit using Markov Chain Monte Carlo [24]; and a new methodology is presented in [25] to compare Kaplan-Meier curves by means of a shiny app. Likewise, multiple shiny apps have been developed for purely practical purposes. For instance, [26,27] for applications in the food field and [28,29] for applications in the industrial sector.

A new open access shiny app for modelling lifetimes through PHD is proposed in the current paper, with no programming skills required to use it. This interactive web application is composed by two blocks: an introductory part where the main features of PHD are explained graphically when different structures are considered for the transition matrix; and a module where the user can estimate the best PHD for a dataset by using multiple estimation options available in the R package *mapfit* [30]. Here, the user can compare the graphic fit of several PHD while also estimating their parameters and assess the goodness of fit. This same methodology is extended in the developed shiny app for the case of one cut-point PHD [31]. This new class of distributions aims to reduce the number of parameters to be estimated and to improve the quality of the fit, especially in the tails of heavy distributions where the classical PHD might provoke an inaccurate fit. Note that for fitting classical probability distributions, R contains a package called *rriskDistributions* [32], in which users can also choose the most appropriate distribution without any knowledge of the R syntax.

Apart from this introduction, the rest of the manuscript is organized as follows: the main theoretical concepts related to PHD and one cut-point PHD are described in Section 2. Section 3 contains a detailed explanation of the developed shiny app. An illustrative example about how to use the app from a users' perspective is displayed in Section 4. Besides, a real dataset that contains the survival time since the diagnostic in primary breast cancers patients has been modelled in Section 5. Section 6 contains a brief discussion about future improvements on the app and the possibilities of using PHD on Matlab and Python. Conclusions are given in Section 7.

## 2. Materials and methods

In this Section, the main theoretical concepts related with Phase-type distributions and one cut-point Phase-type distributions are detailed.

### 2.1. Phase-type distributions

As it was stated in the introduction, PHD are defined as the distribution of the lifetime up to the absorption in an absorbent Markov chain with finite state space. Intuitively, this means that the process is initially found in some of the states (not necessarily in the first one) in which it stays a certain time up to jumping towards another state. This process is repeated a number of times up to reaching the absorbent state in which the process dies.

Formally, a non-negative random variable $X$ is phase-type distributed with representation $(\boldsymbol{\alpha}, \mathbf{T})$ if its cumulative distribution function is given by the following expression

$$F(x) = P[X \leq x] = 1 - \boldsymbol{\alpha} \exp(\mathbf{T}x)\, \boldsymbol{e}, \quad x \geq 0,$$

where $\boldsymbol{e}$ is a column vector of ones with appropriate order, $\boldsymbol{\alpha} = (\alpha_1, \dots, \alpha_m)$ is the vector that represent the initial distribution of the process with $\alpha_i$ being the probability of being initially in the state $i$ and $\mathbf{T}$ is the matrix (sub-generator of order $m$) that contains the transition intensities among the transient states.

From this definition, other interesting functions employed in the field of reliability analysis can be obtained:

- *Density probability function:*

$$f(x) = \boldsymbol{\alpha} \exp(\mathbf{T}x)\mathbf{T}^0, \ \ x \geq 0,$$

with $\mathbf{T}^0 = -\mathbf{T}\boldsymbol{e}$ being the vector that encloses the transition intensities from a transient state up to the absorbent state. Note that $f(x)dx = F(x + dx) - F(x)$ is the probability of the event occurs in the infinitesimal interval $(x, x + dx)$.

- *Survival function (*also called *reliability function):*

$$R(x) = 1 - F(x) = \boldsymbol{\alpha} \exp(\mathbf{T}x)\boldsymbol{e}, \ \ x \geq 0.$$

This measure represents the probability that the system does not fail in the interval $(0, x)$.

- *Hazard rate:*

$$h(x) = \frac{f(x)}{R(x)} = \frac{\boldsymbol{\alpha} \exp(\mathbf{T}x)\mathbf{T}^0}{\boldsymbol{\alpha} \exp(\mathbf{T}x)\boldsymbol{e}}, \ \ x \geq 0.$$

This function is the instantaneous rate of failure and its evolution determines the tendency to the failure in each instant. It is usual that this measure is not decreasing in certain periods of time, since as time goes by the failure tends to increase.

- *Cumulative hazard rate:*

$$H(x) = -\ln\big(1 - F(x)\big) = -\ln(\boldsymbol{\alpha} \exp(\mathbf{T}x)\boldsymbol{e}), \ \ x \geq 0.$$

It can also be computed as the integral of hazard rate. Then, if $H(x)$ evolves in a linear way, the hazard rate is constant (constant failure rate); if the growth is quicker than linear, the function is convex and the hazard rate is increasing (increasing failure rate); if the growth is slower than linear, the function is concave and the hazard rate is decreasing (decreasing failure rate).

- *Mean and variance:*

$$\mathrm{E}[X] = -\boldsymbol{\alpha}\mathbf{T}^{-1}\boldsymbol{e};$$

$$\mathrm{VAR}[X] = 2\boldsymbol{\alpha}\mathbf{T}^{-2}\boldsymbol{e} - (-\boldsymbol{\alpha}\mathbf{T}^{-1}\boldsymbol{e})^2.$$

The real form of all these functions will depend on the structure of the transition intensities matrix, since PHD do not have a unique representation. In fact, PHD characterize a large number of well-known distributions. Some of these are detailed with the corresponding PHD representation:

1. Exponential distribution:

$$F(x) = 1 - \exp(-\lambda x), \ \ x \geq 0 \colon \boldsymbol{\alpha} = 1, \ \mathbf{T} = -\lambda \text{ and } m = 1.$$

2. Erlang distribution: $F(x) = 1 - \sum_{j=0}^{m-1} \exp(-\lambda x)(\lambda x)^j / j!$ for $x \geq 0, m \geq 1$ and $\lambda > 0$,

$$\boldsymbol{\alpha} = (1, \dots, 0); \quad \mathbf{T} = \begin{pmatrix} -\lambda & \lambda & & \\ & -\lambda & \ddots & \\ & & \ddots & \lambda \\ & & & -\lambda \end{pmatrix}_{m \times m}.$$

3. Hypo-exponential distribution: $F(x) = 1 - \sum_{v=0}^{x} \sum_{i=1}^{m} \lambda_i \exp(-\lambda_i v) \left( \prod_{\substack{j=1 \\ j \neq i}}^{m} \frac{\lambda_j}{\lambda_j - \lambda_i} \right)$ for $x \geq 0$ and $\lambda_i \neq \lambda_j$ for $i \neq j$ with $\lambda_i, \lambda_j > 0$,

$$\boldsymbol{\alpha} = (1, \dots, 0); \quad \mathbf{T} = \begin{pmatrix} -\lambda_1 & \lambda_1 & & \\ & -\lambda_2 & \ddots & \\ & & \ddots & \lambda_{m-1} \\ & & & -\lambda_m \end{pmatrix}_{m \times m}.$$

4. Hyper-exponential distribution: $F(x) = 1 - \sum_{i=1}^{m} \alpha_i (1 - \exp(-\lambda_i x))$ for $x \geq 0$ and $\lambda_i > 0$,

$$\boldsymbol{\alpha} = (\alpha_1, \dots, \alpha_m); \quad \mathbf{T} = \begin{pmatrix} -\lambda_1 & & & \\ & -\lambda_2 & & \\ & & \ddots & \\ & & & -\lambda_m \end{pmatrix}_{m \times m}.$$

5. Coxian distribution: for $\lambda_i > 0$ with $i = 1, \dots, m$ and $0 < g_j \leq 1$ with $j = 1, \dots, m-1$,

$$\boldsymbol{\alpha} = (1, \dots, 0); \quad \mathbf{T} = \begin{pmatrix} -\lambda_1 & g_1 \lambda_1 & & \\ & -\lambda_2 & \ddots & \\ & & \ddots & g_{m-1} \lambda_{m-1} \\ & & & -\lambda_m \end{pmatrix}_{m \times m}.$$

6. Generalized Coxian distribution: for $\lambda_i > 0$ with $i = 1, \dots, m$ and $0 < g_j \leq 1$ with $j = 1, \dots, m-1$,

$$\boldsymbol{\alpha} = (\alpha_1, \dots, \alpha_m); \quad \mathbf{T} = \begin{pmatrix} -\lambda_1 & g_1 \lambda_1 & & \\ & -\lambda_2 & \ddots & \\ & & \ddots & g_{m-1} \lambda_{m-1} \\ & & & -\lambda_m \end{pmatrix}_{m \times m}.$$

The estimation of the parameters is not a simple issue because of PHD representation is not unique, which complicates the problem of optimization. In the field of reliability analysis, it is very common to employ a graphical analysis when the estimation process of the parameters by maximum likelihood method presents serious problems of calculus. This parametric graphical technique is based on the principle of least squares and enables a first graphical idea of the fit [33]. However, this method cannot be used in the context of PHD, because PHD cannot be linearized. Consequently, an iterative method called EM-algorithm must be used to estimate the PHD parameters by maximum likelihood [34, 35]. This algorithm is already implemented in most statistical software. An interesting adaption of this method was proposed in [36], in which a partial imputation EM-algorithm was developed for time-to-event modelling of survival data.

## 2.2. One cut-point phase-type distributions

This class of distributions was introduced to improve the quality of the fit and to reduce the number of parameters to be estimated in comparison with the classical PHD, especially in those situations where the distribution presents heavy tails or has two modes [31]. The underlying idea is to determine a suitable cut-point that delimits properly the distribution. Once both zones are defined, two transition intensities matrices are defined to control the behavior in each interval. Given that the transition intensities are different in each zone, the internal performance of the states is not the same over time (non-homogeneous Phase-type distributions).

Formally, a non-negative random variable $Y$ is one cut-point phase-type distributed with representation $(\boldsymbol{\alpha}, \mathbf{T_1}, \mathbf{T_2}, a)$ if its density probability function adopts the following expression

$$f(y) = \begin{cases} \boldsymbol{\alpha}\exp{(\mathbf{T_1}y)}\mathbf{T_1^0} & ; \quad y \leq a \\ \boldsymbol{\alpha}\exp{(\mathbf{T_1}a)}\exp{(\mathbf{T_2}(y-a))}\mathbf{T_2^0} & ; \quad y > a \end{cases},$$

with $a$ being the cut-point, $\mathbf{T_1}$ and $\mathbf{T_2}$ the transition intensities matrix of order $m$ for each zone and $\boldsymbol{\alpha}$ the vector of order $m$ that contains the probabilities of being initially in any state. Note that $\mathbf{T_1}$ and $\mathbf{T_2}$ can have different internal structures.

In a similar way than the classical PHD, the survival function, the hazard rate, the cumulative hazard rate as well as the mean and variance can be derived from the definition.

- *Survival function:*

$$R(y) = \begin{cases} \boldsymbol{\alpha}\exp{(\mathbf{T_1}y)}\boldsymbol{e} & ; \quad y \leq a \\ \boldsymbol{\alpha}\exp(\mathbf{T_1}a)\exp\big(\mathbf{T_2}(y-a)\big)\boldsymbol{e} & ; \quad y > a \end{cases}.$$

- *Hazard rate:*

$$h(y) = \frac{f(y)}{R(y)} = \begin{cases} \dfrac{\boldsymbol{\alpha}\exp{(\mathbf{T_1}y)}\mathbf{T_1^0}}{\boldsymbol{\alpha}\exp{(\mathbf{T_1}y)}\boldsymbol{e}} & ; \quad y \leq a \\ \dfrac{\boldsymbol{\alpha}\exp(\mathbf{T_1}a)\exp{(\mathbf{T_2}(y-a))}\mathbf{T_2^0}}{\boldsymbol{\alpha}\exp(\mathbf{T_1}a)\exp\big(\mathbf{T_2}(y-a)\big)\boldsymbol{e}} & ; \quad y > a \end{cases}.$$

- *Cumulative hazard rate:*

$$H(y) = -\ln\big(1 - F(y)\big) = \begin{cases} -\ln(\boldsymbol{\alpha}\exp(\mathbf{T_1}y)\,\boldsymbol{e}) & ; \quad y \leq a \\ -\ln(\boldsymbol{\alpha}\exp(\mathbf{T_1}a)\exp\big(\mathbf{T_2}(y-a)\big)\boldsymbol{e}) & ; \quad y > a \end{cases}.$$

- *Mean and variance:*

$$\mathrm{E}[Y] = -\boldsymbol{\alpha}\mathbf{T_1^{-1}}\boldsymbol{e} + \boldsymbol{\alpha}\exp(\mathbf{T_1}a)\,(\mathbf{T_1^{-1}} - \mathbf{T_2^{-1}})\boldsymbol{e};$$

$$\mathrm{E}[Y^2] = 2\boldsymbol{\alpha}\mathbf{T}_1^{-2}\boldsymbol{e} - 2\boldsymbol{\alpha}\exp(\mathbf{T}_1 a)[\mathbf{T}_2^{-1}(a\mathbf{I} - \mathbf{T}_2^{-1}) - \mathbf{T}_1^{-1}(a\mathbf{I} - \mathbf{T}_1^{-1})]\,\boldsymbol{e};$$

$$\mathrm{VAR}[Y] = E[Y^2] - (E[Y])^2$$

with $\mathbf{I}$ being the identity matrix of appropriate order.

The probability density function and the hazard rate are discontinuous at point $a$, but the rest of functions do present continuity at this point.

Finally, an interesting result associated with this new class of distributions is that the one cut-point PHD inherits the features of classical PHD. The details about the estimation of parameters can be checked in [31].

## 3. Shiny app

The developed Shiny App aims to provide a new tool to use Phase-type distributions in a friendly environment for those researchers who do not have enough skills of programming and whose area of interest covers survival and reliability analysis. This open access app is placed at a storage cloud belonging to the authors. The instructions to access the app are described in Supplementary.

The Shiny App is composed by four modules at present. Next, these modules are described.

### 3.1. Module 1: Phase-type distributions (PHD)

This part contains the definition for a random variable with Phase-type distribution. Furthermore, it appears the expressions that the transition intensities matrix and the initial distribution vector adopt for different internal structures (i.e. Exponential distribution, Erlang distribution, Coxian distribution, etc.). Finally, once the distribution is chosen, the user can see how $f(x)$, $R(x)$, $h(x)$ and $H(x)$ change graphically when the number of states and the values belonging to $\boldsymbol{\alpha}$ and $\mathbf{T}$ are modified. Figure 1 displays a screenshot of this first module, in which a general distribution with two phases is taken as example. Note that "maximum time of domain" represents the X-axis longitude, that is, the period of observation.

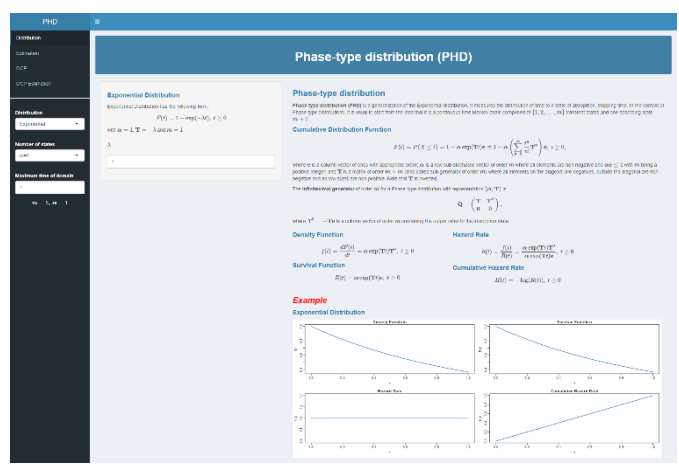

**Figure 1.** Window of shiny app for the first module.

### 3.2. Module 2: Estimation by using Phase-type distributions (PHD)

This module is aimed at modelling an experimental dataset through PHD. Firstly, user must load the file that contains the experimental observations. The app admits files in txt or csv format, while observations must be registered in a single column with the first row being the header. The second step would be selecting the estimation method. The following methods are available in the app:

- *Point data* (recommended): Estimation of the parameters by maximum likelihood from point and weighted point data.
- *Pooled data:* Estimation of the parameters by maximum likelihood from grouped and truncated data [37, 38, 39].
- *Density function:* Estimation of the parameters from a density function defined on the non-negative half-line. This option calls the estimation of *point data* after making weighted point data, which are generated by numerical quadrature.

Finally, once the user chooses the number of states, the app obtains the results automatically (see Figure 2). In particular, the estimation of the parameters for different PHD together with the p-value related to the Anderson-Darling goodness-of-fit test are shown. In addition, both the experimental mean and variance and the associated one with each of these distributions are also computed. The quality of the fitting can be checked graphically by means of the density function, the survival function, the hazard rate and the cumulative hazard rate.

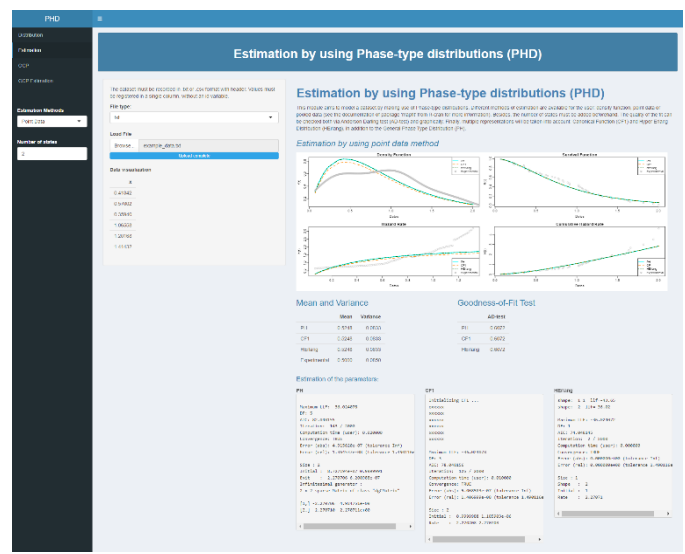

**Figure 2.** Window of shiny app for the second module.

### 3.3. Module 3: One cut-point Phase-type distribution

The definition and the link for the paper where the one cut-point Phase-type distribution was introduced can be seen in this node. Just like in the first module, once the internal structure is selected, user can see how $f(y)$, $R(y)$, $h(y)$ and $H(y)$ change graphically when the number of states and the values belonging to tuple $(\boldsymbol{\alpha}, \mathbf{T_1}, \mathbf{T_2}, a)$ are modified (see Figure 3). Note that in the app current version, $\mathbf{T_1}$ and $\mathbf{T_2}$ are designed with the same internal structure, although it is not a necessary condition from a theoretical viewpoint.

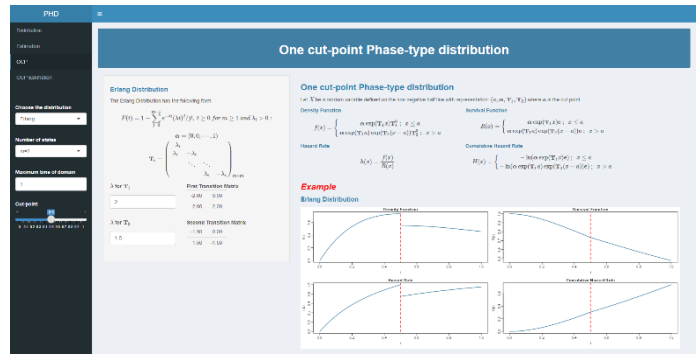

**Figure 3.** Window of shiny app for the third module.

*3.4. Module 4:Estimation by using the one cut-point Phase-type distribution*

This node aims to modelling an experimental dataset through one cut-point PHD. The user must only load the file (in txt or csv format, with the observations registered in a single column with the first row being the header), select the number of states and introduce the cut-point value (by default, the number of state is two and the cut-point value is the mean of the experimental data with two decimals). Subsequently, the Shiny App obtains the outcomes automatically (see Figure 4). Besides, the fitting obtained by the one cut-point PHD is compared with the adjustment reached by the classical PHD. This comparison is made both via goodness-of-fit test and graphically.

To reduce the computational cost in the estimation process, and due to the Erlang distribution has shown a good behavior in the one cut-point PHD, $\mathbf{T}$, $\mathbf{T_1}$ and $\mathbf{T_2}$ have Erlang structure in the app current version.

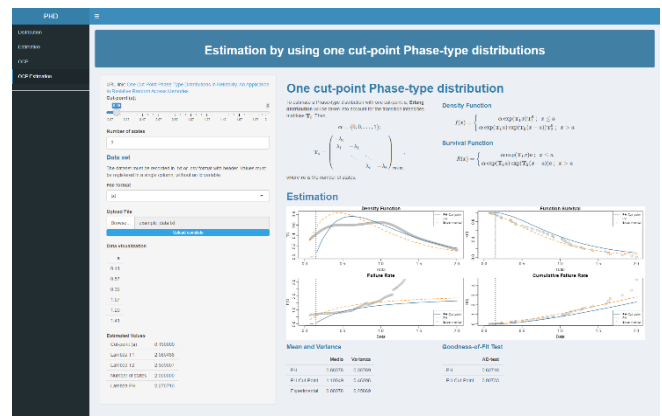

**Figure 4.** Window of shiny app for the fourth module.

## 4. Numerical example with the app

In this Section, a practical example is carried out to improve the description of app and explain how to obtain the results from users' perspective. Given that Module 1 and Module 3 aims to introduce theoretically the classical and one cut-point Phase-type distributions, these modules will not be used hereinafter. In fact, these distributions have already been defined in Section 2.

## 4.1. Preparing the dataset

A dataset with 500 sample points has been simulated from a Weibull distribution with values 3 and 0.5 for shape and scale parameters, respectively. Data has been simulated through the R software, fixing the seed in 5. As it was reported in Section 3, the file that contains the observations must be recorded in .txt or .csv format with header and in a single column (see Figure 5). Note that the decimal separator must be set with points and the header name should be simple and written between quotes.

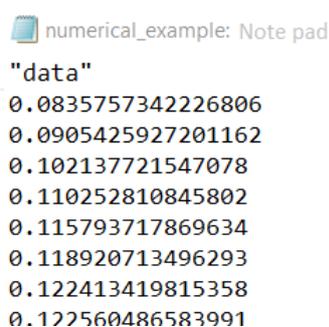

**Figure 5.** Simulated data saved in .txt format with header and in a single column.

## 4.2. Obtaining the results

Let us suppose dataset contains the information about the lifetime of some system. The first step is to test if some classical PHD can be considered to model the dataset. This analysis is done in Module 2. Once the file is loaded and the estimation method is selected (here, the best one is the *point data* method because data are not truncated or grouped), results are obtained automatically for a transition matrix with two internal states (value defined by default). Although app shows multiple measures, firstly users must pay attention to goodness-of-fit tests and the cumulative hazard rate plot. Goodness-of-fit tests are useful to check if the corresponding distribution is suitable to model the dataset, while cumulative hazard rate is highly used in reliability and survival analysis because of its interpretation. Then, the procedure would consists of increasing the number of states up to reach a good fitting. Table 1 and Figure 6 contains the results by considering different internal phases for transition matrix.

**Table 1.** Results of Anderson-Darling goodness-of-fit test for different number of states for each distribution considered.

| Number of states | Distribution | Anderson Darling test (p-value) |
|---|---|---|
| 4 | PH (general structure) | <0.00001 |
| | CF1 (canonical structure) | <0.00001 |
| | HErlang (Hyper-Erlang structure) | <0.00001 |
| 6 | PH (general structure) | 0.0292 |
| | CF1 (canonical structure) | 0.0292 |
| | HErlang (Hyper-Erlang structure) | 0.0257 |
| 8 | PH (general structure) | 0.2044 |
| | CF1 (canonical structure) | 0.1988 |
| | HErlang (Hyper-Erlang structure) | 0.0291 |

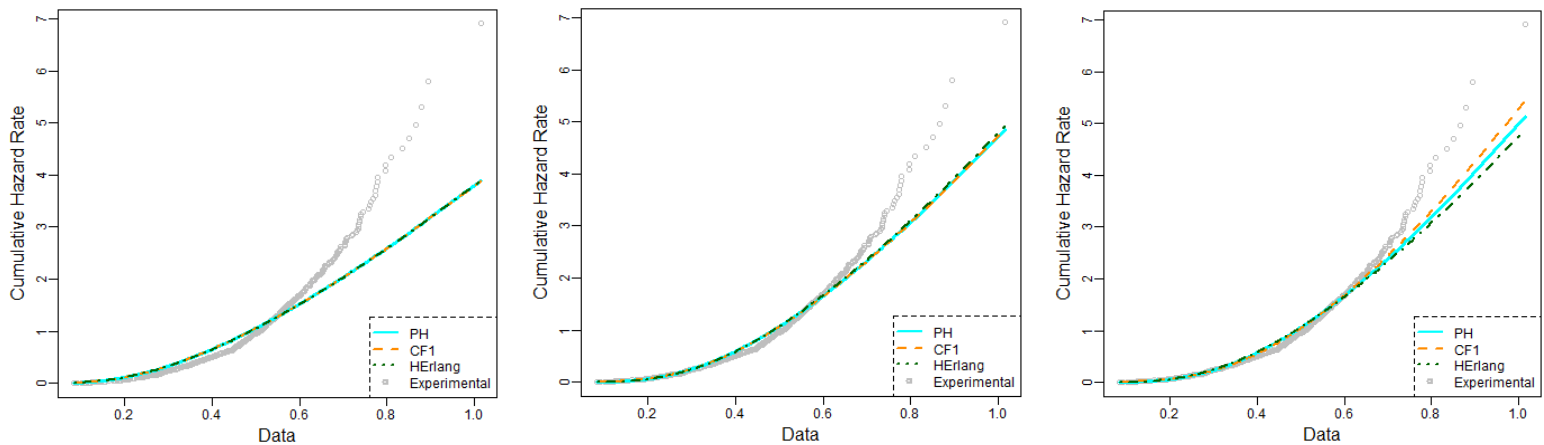

**Figure 6.** Experimental cumulative hazard rate and the corresponding fits. The fit with four states is in top panel, with six states in the central panel and with eight states in the bottom panel.

As it was expected, the quality of the fit is better as the number of states increases. In fact, all distributions are rejected when four internal states are considered but they are accepted for the case of six phases whether the significance level is established in 0.01 (all p-values are greater than the significance level). On the opposite, if the significance level is 0.05 (a common choice for statistical tests), eight phases are required for the case of a distribution with canonical structure or general structure (distribution with hyper-Erlang structure would need more states to be accepted). However, as it is reported in the cumulative hazard rate plots, the improvement with eight phases is not significant in comparison with the fit through six states. Therefore, six internal states is the optimum to avoid increasing the number of parameters to be estimated in the optimization process unnecessarily.

Next, one cut-point PHD is adjusted in order to overcome the fits obtained in Module 2. This analysis is carried out in Module 4, in which data must be loaded again. Again, app shows systematically the results (by default the cut-point is the mean of experimental data and the number of internal states for the transition matrices is two). Remember that the fit via one cut-point PHD is compared with the adjustment reached by a classical PHD with Erlang structure in this Module. An advisable practice is to start the analysis fixing the number of states obtained in Module 2, as the number of phases required in the one cut-point PHD is always less or equal. Once the number of phases is established, detecting a change of trend in the cumulative hazard rate helps to enclose the range of values for the cut-point (users must follow a trial-and-error policy in the current version of app). In particular, the cut-point is settled in 0.58 for this numerical example. As it can be seen in Table 2 and Figure 7, the quality of the fit improves significantly when one cut-point PHD is considered.

**Table 2.** Results of Anderson-Darling goodness-of-fit test for each distribution considered.

| Number of states | Distribution | Anderson Darling test (p-value) |
|---|---|---|
| 6 | PHD (Erlang structure) | 0.0257 |
| | One cut-point PHD | 0.2442 |

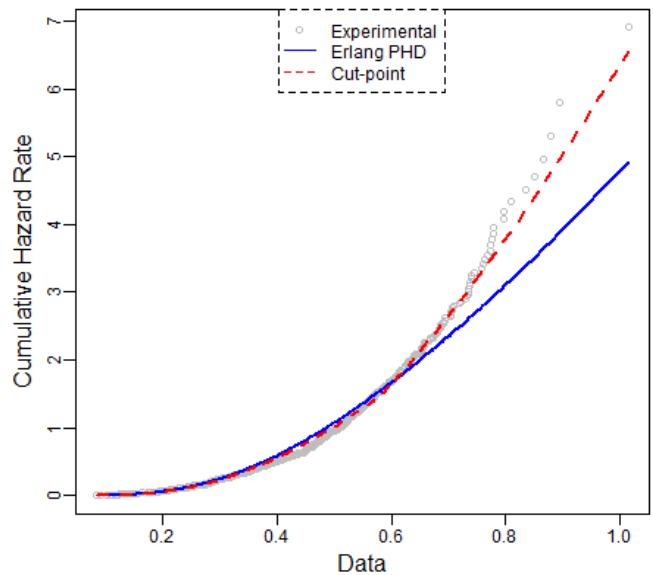

**Figure 7.** Experimental cumulative hazard rate and the corresponding fits.

Finally, several attempts might be done by reducing the number of states with the purpose of finding a more parsimonious model. However, the value of cut-point may be altered slightly as consequence of modifying the dimension of transition intensities matrices.

## 5. Breast cancer dataset

This public dataset contains the information of 2982 primary breast cancers patients whose records were incorporated in the Rotterdam tumor bank [40]. This dataset can be found in the R package *survival*. In the current paper, the time elapsed (in days) since the diagnostic up to the death in those patients who had a recurrence of the illness is analyzed, being the final sample size equals to 1077.

Firstly, different classical probability distributions such as Lognormal, Weibull or Cauchy, among others, were considered without success (p-value associated with the Anderson-Darling goodness-of-fit test is less than 0.05 in all cases). Then, an approach based on PHD is suggested. On the one hand, a Phase-type distribution with general structure was estimated. After the analysis, the optimum value is reached for 3 state with the following representation

$$\boldsymbol{\alpha} = (0,0,1); \ \mathbf{T} = \begin{pmatrix} -0.001254913 & 0 & 0 \\ 0.006190975 & -0.007093341 & 0.0009023664 \\ 0 & 0.001480004 & -0.001483781 \end{pmatrix}.$$

On the other hand, a Phase-type distribution with Erlang structure was considered as well. In this case, the optimum number of phases is two with $\hat{\lambda} = 0.00116$. Finally, a one cut-point PHD was adjusted, in which an Erlang structure was also assumed for the transition intensities matrices. In particular, a one cut-point PHD with the following representation $(\boldsymbol{\alpha}, \mathbf{T_1}, \mathbf{T_2}, a)$ is estimated

$$\boldsymbol{\alpha} = (1,0); \mathbf{T_1} = \begin{pmatrix} -0.00113 & 0.00113 \\ 0 & -0.00113 \end{pmatrix}; \ \mathbf{T_2} = \begin{pmatrix} -0.00175 & 0.00175 \\ 0 & -0.00175 \end{pmatrix}; \ a = 3250.$$

Table 3 contains the goodness-of-fit of each of these distributions. Only the PHD with general structure and the one cut-point PHD must not be rejected (p-values greater than 0.05). However, the best option is the one cut-point PHD (see Figure 8 and the Akaike Information Criterion (AIC) value in Table 3). This choice is supported by the fact that the general PHD does not have a good behavior in the distribution tail (the fitting is not enough accurate in this region). Consequently, relying on the general PHD might lead to ignoring information from patients who lived more than ten years (approximately), as the classical PHD shows a poor fitting around 3600 days.

In particular, a distinct internal behavior in the illness is observed. That is, there is a non-homogeneity in the performance of the illness before and after the cut-point. In both cases, the illness goes through two evolutionary internal states in which it stays around 894 days (on average) in the first phase and approximately 571 days in the second phase. Therefore, the mean time is reduced roughly 35.43% in the second part. This means that the evolution of the illness up to the death is hastened from 3250 days. In conclusion, the model detects two hidden internal states in the behavior of the illness evolution and moreover, the velocity of change in the illness is accelerated from the cut-point, so that the sojourn mean time decreases in each state.

**Table 3.** Summary for the PH and one cut-point adjustments for dataset.

| Distribution | Anderson Darling test (p-value) | logL | AIC value |
|---|---|---|---|
| General PHD | 0.3439 | -8912.38 | 17846.75 |
| Erlang PHD | 0.02395 | -8925.59 | 17857.17 |
| One cut-point PHD | 0.1131 | -8915.34 | 17834.67 |

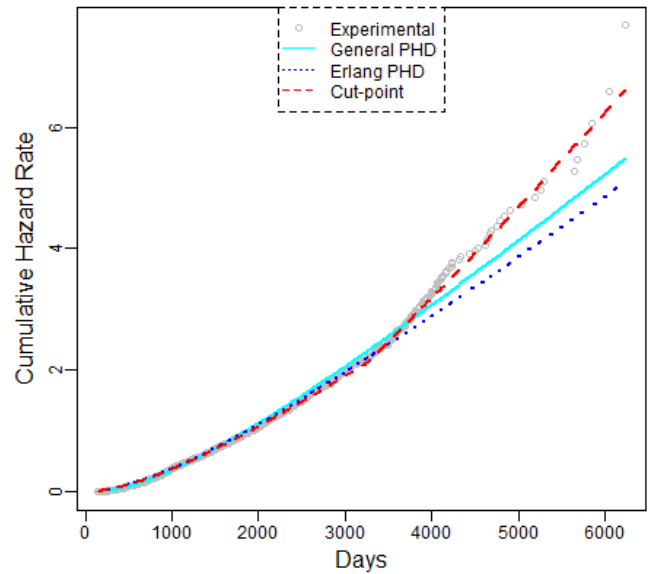

**Figure 8.** Experimental cumulative hazard rate and the corresponding fits.

## 6. Discussion

The current app contains a brief introduction, in a theoretical way, where users can graphically see how the main functions (density function, survival function, etc.) associated with a random variable phase-type distributed change when its parameters are modified. On the other hand, users can also model the lifetime of any system by adjusting some Phase-type distribution. In this panel, users can choose multiple estimation methodologies and compare the quality of the fit both via goodness-of-fit tests and graphically. All these options are extended for the case of one cut-point PHD.

While it is true that this app can already carry out complex statistical analysis, authors have in mind to introduce some improvements in the following versions. Some of the most pressing would be including a module where different classical probability distributions might be adjusted and adding the multiple cut-points Phase-type distributions [41]. In the background, inserting more options for the transition intensities matrices of one cut-point PHD or incorporating an estimation method based on moments (see, for example, [42, 43]) would be also interesting. Likewise, reducing the computational cost in the estimation process of parameters is a big challenge for the future as well.

Finally, as complementary material for those readers who are familiarized with multiple programming environments, some packages available on R, Matlab and Python associated with Phase-type distributions are listed next:

**Table 4.** Options available on R, Matlab and Python to handle PHD.

| R | Matlab | Python |
|---|---|---|
| • Package *mapfit* | • Package *BuTools* [44] | |
| • Package *PhaseTypeR* | • [1]Github: mochamadnurq/ | • Package *pyphase* [46] |
| • Function *PhaseType* | reliability-basicsystem- | • Package *BuTools* [44] |
| from Package *actuar* | phasetype | |

[1] This folder contains MATLAB scripts related to the results obtained in [45].

## 7. Conclusions

A new interactive shiny app is presented in the current manuscript to offer the possibility of fitting Phase-type distributions to an experimental dataset with just some clicks. This open access app has been developed as an installable and standalone local desktop application, making our shiny app accessible to users from all technical levels. In particular, the app has been used to model a simulated dataset and the time to live since the diagnostic in primary breast cancers patients. Take into consideration that this app not only is functional in biomedicine and with other types of cancer but also for any kind of illness, as well as in other fields such as survival, reliability, engineering, and electronics, where the goal is to model the lifetime of some systems (lifetime of tires, supplied voltage in which a device breaks, sojourn time as unemployed, etc.). As far as we know, it is the first web application available to deal with Phase-type distributions, so we look forward to it is welcome in any area of science.

## Use of AI tools declaration

The authors declare they have not used Artificial Intelligence (AI) tools in the creation of this article.

## Acknowledgments

Research supported by the project PID2020-113961GB-I00 funded by MCIN/ AEI /10.13039/501100011033 of the Spanish Ministry of Science and Innovation (also supported by the FEDER programme), the "Maria de Maeztu" Excellence Unit IMAG reference CEX2020-001105-M, funded by MCIN/AEI/10.13039/501100011033/ and by the project FQM-307 of the Government of Andalusia (Spain).

## Conflict of interest

The authors declare there is no conflict of interest.

## References

1. F.P. Coolen, Parametric probability distributions in reliability, in *Encyclopedia of Quantitative Risk Analysis and Assessment*, John Wiley & Sons: Chichester, (2008), 1255-1260.
2. J. W. McPherson, *Reliability physics and engineering: time-to-failure modelling*, Springer: Heidelberg, 2013.
3. A. D. Hutson, An Accelerated Life Model Analog for Discrete Survival and Count Data, *Comput. Meth. Prog. Bio.*, **210** (2021), 106337. https://doi.org/10.1016/j.cmpb.2021.106337
4. M.C. Aguilera-Morillo, A.M. Aguilera, F. Jiménez-Molinos, J.B. Roldán, Stochastic modeling of Random Access Memories reset transitions, *Math. Comput. Simulat.*, **159** (2019), 197-209. https://doi.org/10.1016/j.matcom.2018.11.016

5.  R. Kollu, S.R. Rayapudi, S. Narasimham, K.M. Pakkurthi, Mixture probability distribution functions to model wind speed distributions, *Int. J. Energ. Environ. Eng.*, **3** (2012), 27. https://doi.org/10.1186/2251-6832-3-27

6.  F.J. Marques, C.A. Coelho, M. de Carvalho, On the distribution of linear combinations of independent Gumbel random variables, *Stat. Comput.*, **25** (2015), 683-701. https://doi.org/10.1007/s11222-014-9453-5

7.  M. F. Neuts, *Probability distributions of phase type*, Liber Amicorum Prof. Emeritus H. Florin, 1975.

8.  M.F. Neuts, *Matrix-Geometric Solutions in Stochastic Models: An Algorithmic Approach*, John Hopkins University Press: Baltimore, 1981.

9.  M. Kijima, *Markov processes for stochastic modelling*, Springer: New York, 2013.

10. V.G. Kulkarni, *Modeling and analysis of stochastic systems*, Crc Press, 2016.

11. Q.M. He, *Fundamentals of matrix-analytic methods*, Springer: New York, 2014.

12. S. Asmussen, *Ruin probabilities*, World Scientific, 2000.

13. S. Mahmoodi, S. Hamed Ranjkesh, Y.Q. Zhao, Condition-based maintenance policies for a multi-unit deteriorating system subject to shocks in a semi-Markov operating environment, *Qual. Eng.,* **32** (2020), 286-297. https://doi.org/10.1080/08982112.2020.1731754

14. E. Pérez, D. Maldonado, C. Acal, J.E. Ruiz-Castro, A.M. Aguilera, F. Jiménez-Molinos, J.B. Roldán, C. Wenger, Advanced temperature dependent statistical analysis of forming voltage distributions for three different HfO2-based RRAM technologies, *Solid State Electron.*, 176 **(2021)**, 107961. https://doi.org/10.1016/j.sse.2021.107961

15. J.E. Ruiz-Castro, C. Acal, A.M. Aguilera, J.B. Roldán, A complex model via phase-type distributions to study random telegraph noise in resistive memories, Mathematics, **9** (2021), 390. https://doi.org/10.3390/math9040390

16. S. Gordon, A.H. Marshall, M. Zenga, Predicting elderly patient length of stay in hospital and community care using a series of conditional coxian phase-type distributions, further conditioned on a survival tree, *Health Care Manag. Sc.*, **21** (2018), 269-280. https://doi.org/10.1007/s10729-017-9411-9

17. M. Bladt, A review on phase-type distributions and their use in risk theory, *ASTIN Bull.: The Journal of the IAA*, **35** (2005), 145-161. https://doi.org/10.1017/s0515036100014100

18. J.E. Ruiz-Castro, C. Acal, A.M. Aguilera, M.C. Aguilera-Morillo, J.B. Roldán, Linear-Phase-Type probability modelling of functional PCA with applications to resistive memories, *Math. Comput. Simulat.*, **186** (2021), 71-79. https://doi.org/10.1016/j.matcom.2020.07.006

19. W. Chang, J. Cheng, J.J. Allaire, C. Sievert, B. Schloerke, Y. Xie, J. Allen, J. McPherson, A. Dipert, B. Borges, R package shiny (2022). Available from: https://CRAN.R-project.org/package=shiny

20. M.G. Genton, S. Castruccio, P. Crippa, S. Dutta, R. Huser, Y. Sun, S. Vettori, Visuanimation in statistics, *Stat*, **4** (2015), 81–96. https://doi.org/10.1002/sta4.77

21. J. Wrobel, S.Y. Park, A.M. Staicu, J. Goldsmith, Interactive graphics for functional data analyses, *Stat*, **5** (2016), 108–118. https://doi.org/10.1002/sta4.109

22. J.P. Fortin, E. Fertig, K. Hansen, shinyMethyl: interactive quality control of Illumina 450k DNA methylation arrays in R, *F1000research*, **3** (2014) 175. https://doi.org/10.12688/f1000research.4680.2

23. C. Tebé, J. Valls, P. Satorra, A. Tobías, COVID19-world: a shiny application to perform comprehensive country-specific data visualization for SARS-CoV-2 epidemic, *BMC Med. Res. Methodol.,* **20** (2020), 235. https://doi.org/10.1186/s12874-020-01121-9

24. J. Gabry, D. Veen, M. Andreae. M. Betancourt, B. Carpenter, Y. Gao, A. Gelman, B. Goodrich, D. Lee, D. Song, R. Trangucci, R package shinystan: Interactive Visual and Numerical Diagnostics and Posterior Analysis for Bayesian Models, (2015). Available from: https://CRAN.R-project.org/package=shinystan

25. N.T. Stevens, L. Lu, Comparing Kaplan-Meier curves with the probability of agreement, *Stat. Med.,* **39** (2020), 4621-4635. https://doi.org/10.1002/sim.8744

26. T.C. Wang, Developing a flexible and efficient dual sampling system for food quality and safety validation, *Food Control,* **145** (2023), 109483. https://doi.org/10.1016/j.foodcont.2022.109483

27. T.C. Wang, Generalized variable quick-switch sampling as a novel method for improving sampling efficiency of food products, *Food Control,* **135** (2022), 108841. https://doi.org/10.1016/j.foodcont.2022.108841

28. M.H. Shu, T.C. Wang, B.M. Hsu, Integrated green-and-quality inspection schemes for green product quality with six-sigma yield assurance and risk management, *Qual. Reliab. Eng. Int.,* **39** (2023), 2720-2735. https://doi.org/10.1002/qre.3381

29. T.C. Wang, B.M. Hsu, M.H. Shu, Quick-switch inspection scheme based on the overall process capability index for modern industrial web-based processing environment, *Appl. Stoch. Model. Bus.,* **38** (2022), 847-861. https://doi.org/10.1002/asmb.2667

30. H. Okamura, T. Dohi, mapfit: An R-Based Tool for PH/MAP Parameter Estimation, in *Quantitative Evaluation of Systems*. QEST 2015. Lecture Notes in Computer Science (vol. 9259), Springer, (2015), 105-112. https://doi.org/10.1007/978-3-319-22264-6_7

31. C. Acal, J.E. Ruiz-Castro, D. Maldonado, J.B. Roldán, One Cut-Point Phase-Type Distributions in Reliability. An Application to Resistive Random Access Memories, Mathematics, **9** (2021), 2734. https://doi.org/10.3390/math9212734

32. N. Belgorodski, M. Greiner, K. Tolksdorf, K. Schueller, R package rriskDistributions: Fitting Distributions to Given Data or Known Quantiles, R package version, (2017). https://CRAN.R-project.org/package=rriskDistributions

33. J.F. Lawless, *Statistical models and methods for lifetime data* (2º ed.), John Wiley & Sons, 2003.

34. S. Asmussen, O. Nerman, M. Olsson, Fitting Phase-Type Distributions via the EM Algorithm, *Scand. J. Stat.*, **23** (1996), 419-441. http://www.jstor.org/stable/4616418.

35. P. Buchholz, J. Kriege, I. Felko, *Input Modeling with Phase-Type Distributions and Markov Models. Theory and Applications*, Cham: Springer, 2014. https://doi.org/10.1007/978-3-319-06674-5.

36. K. Choi, S.M. Park, S. Han, D.S. Yim, A partial imputation EM-algorithm to adjust the overestimated shape parameter of the Weibull distribution fitted to the clinical time-to-event data, *Comput. Meth. Prog. Bio.*, **197** (2020), 105697. https://doi.org/10.1016/j.cmpb.2020.105697

37. A. Thummler, P. Buchholz, M. Telek, A novel approach for phase-type fitting with the EM algorithm, *IEEE T. Depend. Secure,* **3** (2006), 245-258.

38. A. Panchenko, A. Thummler, Efficient phase-type fitting with aggregated traffic traces, *Perform. Evaluation*, **64** (2007), 629-645. https://doi.org/10.1016/j.peva.2006.09.002

39. H. Okamura, T. Dohi, K.S. Trivedi, Improvement of EM algorithm for phase-type distributions with grouped and truncated data, *Appl. Stoch. Model. Bus.*, **29** (2013), 141-156. https://doi.org/10.1002/asmb.1919

40. P. Royston, D.G. Altman, External validation of a Cox prognostic model: principles and methods, *BMC Med. Res. Methodol.*, **13** (2013), 1-15. https://doi.org/10.1186/1471-2288-13-33

41. J.E. Ruiz-Castro, C. Acal, J.B. Roldán, An approach to non-homogenous Phase-type distributions through multiple cut-points, *Qual. Eng.,* **35**(4) (2023), 619-638.

42. A. Bobbio, A. Horvath, M. Telek, Matching three moments with minimal acyclic phase type distributions, *Stoch. Models*, **21** (2005), 303-326. https://doi.org/10.1081/STM-200056210

43. T. Osogami, M. Harchol-Balter, Closed form solutions for mapping general distributions to minimal PH distributions, *Perform. Evaluation*, **63** (2006), 524-552. https://doi.org/10.1016/j.peva.2005.06.002

44. G. Horváth, M. Telek, Markovian Performance Evaluation with BuTools, in *Systems Modeling: Methodologies and Tools*, Springer, Cham, 2019. https://doi.org/10.1007/978-3-319-92378-9_16

45. A. Alkaff, M.N. Qomarudin, Modeling and analysis of system reliability using phase-type distribution closure properties, *Appl. Stoch. Model. Bus.*, **36** (2020), 548-569. https://doi.org/10.1002/asmb.2509

46. M. Langer, Y. Zhang, D. Figueirinhas, J.B. Forien, K. Mom, C. Mouton, R. Mokso, P. Villanueva-Perez, PyPhase–a Python package for X-ray phase imaging, *J. Synchrotron Radiat.*, **28** (2021), 1261-1266. https://doi.org/10.1107/S1600577521004951

**Supplementary**

To open the shiny app, the following steps are required:

1. Download the PHD-Desktop-APP.zip hosted at the following URL:
   https://www.ugr.es/local/jeloy/PHD-Desktop-App.zip

2. Once the file is downloaded, the zip must be decompressed. Users must run the file called "phd.bat" in the unzipped folder. After this step, the shiny app will be open in the browser.

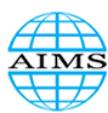